\documentclass[12pt]{iopart}

\pdfoutput=1
\expandafter\let\csname equation*\endcsname\relax
\expandafter\let\csname endequation*\endcsname\relax
\usepackage[english]{babel}
\usepackage{amsmath,
amssymb,amsfonts,latexsym}
\usepackage{graphicx}
\usepackage{mathrsfs} 
\usepackage{color}
\usepackage{booktabs}
\usepackage{cite}
\usepackage{bm}
\usepackage{braket}
\usepackage{nicefrac}

\usepackage{soul}
\usepackage{mathtools}
\usepackage{ulem}
\usepackage{dsfont}

\definecolor{Blue}{rgb}{0.00, 0.00, 1.00}
\definecolor{Red}{rgb}{1.00, 0.00, 0.00}
\definecolor{labelkey}{cmyk}{.1,.7,0.5,0}

\definecolor{mygreen}{rgb}{0,0.5,0}

\definecolor{blue(pigment)}{rgb}{0.2, 0.2, 0.6}
\usepackage{times}
\usepackage[colorlinks,%bookmarks=false,
citecolor=blue,linkcolor=blue,urlcolor=blue]{hyperref}
\hypersetup{linktocpage}
\usepackage{scalerel}

\makeatletter
\def\@mkboth#1#2{}
\newlength\appendixwidth
\preto\appendix{\addtocontents{toc}{\protect\patchl@section}}
\newcommand{\patchl@section}{%
  \settowidth{\appendixwidth}{\textbf{Appendix }}%
  \addtolength{\appendixwidth}{1.5em}%
  \patchcmd{\l@section}{1.5em}{\appendixwidth}{}{\ddt}%
}
\makeatother

\def\eqref#1{(\ref{#1})}
\usepackage{cancel}

\newcommand{\bea}{\begin{eqnarray}}
\newcommand{\eea}{\end{eqnarray}}
\newcommand{\sgn}{\mathrm{sgn}}

\renewcommand*{\geq}{\geqslant}
\renewcommand*{\leq}{\leqslant}
\newcommand{\I}{\ensuremath{\mathbf{i}}}
\newcommand*{\ddt}[1]{%
  \accentset{\mbox{\bfseries .\hspace{-0.25ex}.}}{#1}} % 2 time derivatives 

\newcommand{\be}{\begin{equation}}
\newcommand{\ee}{\end{equation}}

\newcommand{\beA}{\begin{align}}
\newcommand{\eeA}{\end{align}}

\newcommand{\bV}{\begin{pmatrix}}
\newcommand{\eV}{\end{pmatrix}}

\newcommand{\dd}{\mathrm{d}}
\newcommand{\de}{\partial}

\newcommand{\Ha}{\hat{H}} % Notation for the Hamiltonian

\newcommand{\ssz}{\hat{\sigma}^z} % Notation for spin-chain operators
\newcommand{\ssx}{\hat{\sigma}^x}
\newcommand{\ssy}{\hat{\sigma}^y}
\newcommand{\ssp}{\hat{\sigma}^+}
\newcommand{\ssm}{\hat{\sigma}^-}

\newcommand{\Cc}{\hat{c}}

\newcommand{\rrho}{\hat{\rho}}

\graphicspath{{Figures/}}

\begin{document}
\title[]{Exact hydrodynamic description of symmetry-resolved R\'enyi entropies  after a quantum quench}
\author{Stefano Scopa$^{1,\dagger}$ and D\'avid X. Horv\'ath$^{1,*}$}
\address{$^1$ SISSA and INFN, via Bonomea 265, 34136 Trieste, Italy}
\address{email: $\dagger$~\href{mailto:sscopa@sissa.it}{sscopa@sissa.it}; \;
 $*$~\href{mailto:dahorva@sissa.it}{dahorva@sissa.it}
}
\date{\today}
\begin{abstract}
We investigate the non-equilibrium dynamics of the symmetry-resolved R\'enyi entropies in a one-dimensional gas of non-interacting spinless fermions by means of quantum generalised hydrodynamics, which recently allowed to obtain very accurate results for the total entanglement in inhomogeneous quench settings. Although our discussion is valid for any quench setting accessible with quantum generalised hydrodynamics, we focus on the case of a quantum gas initially prepared in a bipartite fashion and subsequently let evolve unitarily with a hopping Hamiltonian. For this system, we characterise the symmetry-resolved R\'enyi entropies as function of time $t$ and of the entangling position $x$ along the inhomogeneous profile. We observe an asymptotic logarithmic growth of the charged moments at half system and an asymptotic restoration of equipartition of entropy among symmetry sectors with deviations which are proportional to the square of the inverse of the total entropy.
\end{abstract}
\hrulefill
{\small \tableofcontents}
\hrulefill
\maketitle

%\newpage
\section{Introduction}
Since the birth of quantum mechanics, the concept of entanglement has been at the core of any quantum theory. Its profound and sometimes subtle links to various aspects of physics, ranging from the connection to thermal entropy \cite{Takahashi-book,leshouches-PC} to the applications in the early universe e.g.~\cite{Boyanovsky2018,Brahma2020,Martin2021}, have been stimulating an enormous scientific activity in the last years. Nowadays, entanglement measures have become a commonly recognised (and very efficient) tool for the investigation and the understanding of quantum correlations. This is particularly true in low-dimensional quantum systems, which are notable for hosting strong quantum correlations,  see e.g. \cite{vNE1,vNE2,vNE3,vNE4} for recent reviews and e.g. \cite{Islam2015,Kaufman2016,Elben2018,Brydges2019,Lukin2019} for some experimental tests.
Alongside a still fascinating research on entanglement in exotic and/or out-of-equilibrium contexts, it has been initiated to formulate refined tools to go beyond the conventional entanglement measures and, in this way, obtain more information on such quantum correlations. An example is the so-called entanglement Hamiltonian e.g. \cite{Li2008,Eisler2017,ep-18,Eisler2019,Cardy2016,wrl-18,Hislop1982,Dalmonte:2017bzm,ksz-21,kbe-21,zksz-22,Javerzat2021} (and more recently the negativity Hamiltonian \cite{Murciano2022}), which encodes in a single object the full description on the entanglement spectrum and on its topological properties, or the understanding of the structure of entanglement with respect to an internal symmetry of the model under analysis \cite{Lukin2019}, which is the main character of this work. Despite the enormous success that the idea of the symmetry resolution of entanglement is experiencing, it was first applied to a concrete physical system only recently in Ref.~\cite{lr-14} and put forward in a more general context even later in Ref.~\cite{gs-18}. %The main target of these studies is essentially to understand how entanglement is structured with respect to a certain internal symmetry of a particular state.
 Typically, one considers models having $U(1)$ internal symmetry associated  with the conservation of the particle number $\braket{\hat{N}}$, but the case of non-abelian symmetries has been also investigated, see Ref.~\cite{cdmWZW-21}.  Here, we focus on the usual case of an abelian internal symmetry by considering a pure quantum state $\vert \Psi \rangle$
such that
\be
[\hat{\rho},\hat{N}]=0\,,\;\;\hat{\rho}=\vert\Psi\rangle\langle\Psi\vert\,.
\ee
Since the particle number operator is made out of the sum of local densities, i.e., $\hat{N}=\sum_{j\in\mathbb{Z}} \hat{n}_j$,  by taking any spatial bi-partition $A\cup \bar{A}\equiv[-\infty,\ell]\cup[\ell+1,\infty]$ of the system with a cut at a certain position $\ell\in\mathbb{Z}$, one finds that
\be
\hat{N}=\hat{N}_A \otimes  \mathds{1}_{\bar{A}}+ \mathds{1}_{A} \otimes \hat{N}_{\bar{A}},
\ee
where $\hat{N}_A=\sum_{j\leq\ell} \hat{n}_j$ and similarly for $\hat{N}_{\bar{A}}$.  It is then easy to show \cite{cdmWZW-21} that the reduced density matrix  $\rrho_A={\rm tr}_{\bar{A}} (\rrho)$, still commutes with $\hat{N}_A$, that is,   
\be
[\hat\rho_A,\hat{N}_A]=0\,.
\label{RDMCommutation}
\ee
Moreover, it is evident that since $\hat{N}$ is a global conserved quantity $[\hat{N},\hat{H}]=0$, Eq.~\eqref{RDMCommutation} remains valid during the course of time evolution as well, i.e.,
\be
[\hat\rho_A(t),\hat{N}_A]=0\,.
\label{RDMCommutationTimeEvol}
\ee

The commutation relations \eqref{RDMCommutation} and \eqref{RDMCommutationTimeEvol} imply a block-diagonal structure of $\hat\rho_A$ in terms of the eigenvalues $N$ of $\hat{N}_A$ as 
\be
\label{rhoN}
\hat\rho_A=\bigoplus_N \hat\Pi_N \ \hat\rho_A = \bigoplus_N \left[ p(N) \hat\rho_A(N)\right]
\ee
where $p(N)={\rm tr}(\hat\Pi_N \hat\rho_A)$ is the probability of having $N$ particles in the subsystem $A$. The symmetry-resolved R\'enyi entropy is then defined as usual
\be\label{SRRE}
S_{n,N}=\frac{1}{1-n} \log\tr\left(\hat\rho_A(N)^n\right)\,,
\ee
via the symmetry-resolved reduced density matrix $\hat\rho_A(N)$. The calculation of $\hat\rho_A(N)$ is typically very challenging, due to the non-local action of the projector $\hat\Pi_N$ on the subspace with $N$ particles, making the calculation of $S_{n,N}$ almost out-of-reach, if one tries to work it out directly from its definition. Nevertheless, it was noticed in Ref.~\cite{gs-18} that an equivalent (and very convenient) formulation of the problem can be done by focusing on the so-called symmetry resolved charged moments using path-integral formalism or known lattice techniques. These quantities were already introduced before their connection to symmetry-resolved entropies were established \cite{bym-13,cms-13,cnn-16,d-16, d-16b,ssr-17,srrc-19}, and in particular,  are defined as
\be
\label{c}
Z_n(\alpha)=\tr\left[\hat\rho_A^n e^{\I \alpha\hat{N}_A}\right],
\ee
implying that their computation is indeed feasible in the path integral formulation of the replicated model at the price of introducing an additional flux on one of the Riemann sheets. 
Equally importantly, one finds that 
\be
\label{SRPartitionFunction}
{\cal Z}_n(N)=\int_{-\pi}^\pi \frac{\dd\alpha}{2\pi} \ e^{-\I N \alpha} Z_n(\alpha)\equiv \tr\left[\hat\Pi_N\hat\rho_A^n\right]\,,
\ee
 for the Fourier transform of the charged moments, aka the symmetry-resolved partition function, and hence the symmetry-resolved R\'enyi entropy \eqref{SRRE} can be expressed as
\be\label{SRRE2}
S_{n,N}=\frac{1}{1-n} \log \left[\frac{{\cal Z}_n(N)}{({\cal Z}_1(N))^n}\right].
\ee
Immediately after the introduction of these concepts in Refs.~\cite{lr-14, gs-18}, the symmetry resolution of entanglement has been investigated in a large variety of systems, including e.g.~1+1 conformal field theories \cite{lr-14,gs-18,Equipartitioning,SREQuench,bc-20,crc-20,Zn,mbc-21, Chen-21, cc-21, cdmWZW-21,ThermalCorrectionSRECFT,Chen-Negativity}, free  \cite{mdc-20b,U(1)FreeFF}  
and interacting integrable quantum field theories \cite{Z2IsingShg, SGSRE, PottsSRE}  and also holographic settings \cite{znm,Zhao2021,Zhao2022}. These studies are generically carried out in  lattice models \cite{lr-14,Equipartitioning,SREQuench,brc-19,fg-20,FreeF1,FreeF2,mdc-20,ccdm-20,pbc-20,bc-20,HigherDimFermions,mrc-20,FFChriticalChain,Ma2021}, but other systems exhibiting more exotic types of dynamics have been also considered  \cite{trac-20,MBL,MBL2,Topology,Anyons, as-20,QuantumHallSRE}. Moreover, the interest for symmetry-resolved entanglement measures on the theoretical side is accompanied and consolidated by their experimental feasibility, see e.g. Ref.~\cite{ncv-21,vecd-20}. \\

%It follows that the study of symmetry resolved quantities has quickly become a consolidate research avenue with many highly significant and challenging ongoing tasks. 
In addition, the investigation of symmetry-resolution in out-of-equilibrium situations has also been initiated, although so far carried out only for very few cases, see \cite{SREQuench, pbc-20,ncv-21,fg-21, pbc-21, pbc-22,Chen-NegativityDynamics}. The reason for this lack in literature is quite evident as the study of out-of-equilibrium entanglement is known to be often challenging and its possible symmetry-resolved counterpart is seen to be even harder to analyse. However, besides being interesting in its own right, probing the non-equilibrium properties of the symmetry-resolved entanglement could significantly help us thoroughly understand these quantities and the eventual physical systems they are associated with. With this scope, in this manuscript, we wish to connect the idea of symmetry resolution with the framework of quantum generalised hydrodynamics \cite{Ruggiero2019,Ruggiero2020,Collura2020,StefanoJerome,Scopa2021b,Rottoli2022,Ruggiero2022,Ares2022}, which recently enabled to obtain very accurate predictions for the non-equilibrium dynamics of the total entropy in non-homogeneous quench settings. \\
More precisely, the aim of this work is twofold: on the one hand, we detail the exact asymptotic solution for a prototypical setting of  non-homogeneous and out-of-equilibrium system, that is, a bi-partitioning quench protocol made with a one-dimensional gas of free spinless fermions, see e.g. Ref.~\cite{Dubail2017,StefanoJerome,Scopa2021b,Rottoli2022,Ares2022} and Sec.~\ref{sec:intro} below. To our best knowledge there are no similar studies about the symmetry-resolved entanglement of such inhomogeneous quench protocols. On the other hand, and most importantly, our discussion on symmetry resolution has a general validity and applies to any inhomogeneous quench protocol which is accessible by quantum generalised hydrodynamics.\\

{\it Outline}~---~In Sec.~\ref{sec:intro}, we briefly introduce the model and the quench protocol considered in this work. Similarly, Sec.~\ref{SemiClassicalHydroDescription} is a short introduction to phase-space hydrodynamics, made by following the discussions in recent works (e.g. \cite{Dubail2017,Collura2020,StefanoJerome,Scopa2021b,Rottoli2022,Ares2022}) and that of previous studies on the quasi-classical evolution of the conserved charges (e.g. \cite{Antal1999,Antal2008,Karevski2002, Rigol2004,Platini2005,Platini2007,Eisler2008,Alba2014,Rigol2015,Allegra2016}). After preparing this ground, in Sec.~\ref{sec:QGHD} we re-quantise the hydrodynamic solution at low energy in terms of a Luttinger liquid by following the recent literature on quantum generalised hydrodynamics, see e.g. \cite{Ruggiero2022,StefanoJerome}. In Sec.~\ref{sec:charge-mom} we specialise to the calculation of symmetry resolved quantities and, in particular, we detail the strategy of calculation of the charged moments in the quantum generalised hydrodynamic framework. Finally, Sec.~\ref{sec:SRRE} contains the analysis of the symmetry-resolved partition function \eqref{SRPartitionFunction} and of the symmetry-resolved R\'enyi entropy \eqref{SRRE2} while Sec.~\ref{sec:conclusions} summarise our work and our results. We provide a numerical check of our major results based on exact lattice calculations, whose implementation details are reported in \ref{app:NUM}.

%%%%%%%%%%%%%%%%%%%%%%%%%%%%%%%%%%%
\section{The model, the quench and the hydrodynamic descriptions}
\subsection{Quantum model and quench protocol}\label{sec:intro}
In this work, we consider a one-dimensional gas of non-interacting spinless fermions on a semi-infinite lattice $j\in [-L,\infty]$ with nearest-neighbour hopping, whose Hamiltonian reads
\be\label{model}
\Ha=-\frac{1}{2}\sum_{j=-L}^\infty\left(\Cc^\dagger_j\Cc_{j+1} + \Cc^\dagger_{j+1}\Cc_j \right),
\ee
where $\Cc^\dagger_j$, $\Cc_j$ are standard fermionic lattice operators satisfying canonical anti-commutation relations $\{ \Cc_i , \Cc_j^\dagger  \} = \delta_{ij}$. The system is initially prepared in a state $\left| \Omega \right>$ obtained as ground state of the trapped Hamiltonian
\be
\label{free-gas}
\Ha_{t<0}=-\frac{1}{2}\sum_{j=-L}^\infty \left(\Cc^\dagger_j\Cc_{j+1} + \Cc^\dagger_{j+1}\Cc_j +(V_j-\mu)  \Cc^\dagger_j\Cc_{j} \right)\,,
\ee
where $\mu$ is a chemical potential and $V_j$ is a confining potential specified as
\begin{equation}\label{potential}
	V_j \, = \,  \left\{ \begin{array}{rcl}
		0  & \text{if  \quad $-L\leq j \leq -1$}; \\
		+\infty  & \text{otherwise}.
	\end{array} \right.
\end{equation}
This setup can be equivalently interpreted as a quench protocol where the trap in Eq.~\eqref{free-gas} is suddenly released at $t=0$ and the model is subsequently evolved with Hamiltonian \eqref{model}. Notice that if we set $\mu=0$ in  Eq.~\eqref{free-gas}, the ground state contains exactly $L/2$ particles (we assume $L$ is even). This can be easily seen by diagonalising the Hamiltonian~\eqref{free-gas} with the potential \eqref{potential} yielding
\be
\Ha_{t<0} =- \sum_k   \cos(k)\;  \hat\eta_k^\dagger \hat\eta_k,
\ee 
with Fourier modes of momentum $k= \pi q/(L+1)$, $q=1,\dots, L$, given as
\be
\hat\eta_k^\dagger= \sqrt{\frac{2}{L+1}} \sum_{j=-L}^{-1}  \sin \left( k j \right) \Cc_j^\dagger,\qquad  \{ \hat{\eta}_k^\dagger, \hat{\eta}_{k'} \}=\delta_{kk'}.
\ee
Indeed, the single-particle energy $-\cos(k)$ is negative for $q=1, \dots,L/2$ and the ground state is obtained by acting on the fermion vacuum $ \left| 0 \right>$ with such single-particle creation operators
\be\label{half-filled-GS}
\ket{\Omega}\equiv \left| \{\rho=1/2\} \right>=\hat{\eta}_1^\dagger  \hat{\eta}_2^\dagger \dots \hat{\eta}_{\nicefrac{L}{2}}^\dagger  \left| 0 \right>.
\ee
Here, we introduced the notation $ \left| \{\rho=1/2\} \right>$ to emphasise that the ground state at $\mu=0$ is half-filled, that is, on average, every second site is occupied by a fermion. Together with the state in Eq.~\eqref{half-filled-GS}, we consider also the case of a fully-filled ground state, obtained by setting $\mu<-1$ in Eq.~\eqref{free-gas} as
\be
\ket{\Omega}\equiv \left| \{\rho=1\} \right>=\hat{\eta}_1^\dagger  \hat{\eta}_2^\dagger \dots \hat{\eta}_{L}^\dagger  \left| 0 \right>.
\ee
Both the variants of the initial states allow for an intuitive spin chain interpretation since the model in Eq.~\eqref{free-gas} is  known to map to a spin-$1/2$ XX-chain
\be\label{xx-model}
\Ha= - \frac{1}{4} \sum_{j=-L}^{\infty} \left(\ssx_j \ssx_{j+1} + \ssy_j\ssy_{j+1}\right) \, + \, \frac{1}{2}\sum_{j=-L}^{\infty} (V_j-\mu) \ssz_j \, + \, {\rm constant}
\ee
through the Jordan-Wigner transformation \cite{Jordan1928}
\be
\Cc_j^\dagger= \exp\left(\I \pi \sum_{i<j} \ssp_i\ssm_i\right) \ssp_j , 
\ee
where $\hat\sigma^\pm_j=(\ssx_j\pm \I \ssy_j)/2$ and $\hat\sigma_j^{a}$, $a=x,y,z$, are spin-$1/2$ operators acting at site $j$. In particular, the specific choice  $\ket{\Omega}\equiv\left| \{\rho=1\} \right>$ corresponds to the standard domain wall where the left and the right parts of the system display opposite value of magnetisation equal to $+\frac{1}{2}$ and $-\frac{1}{2}$ respectively, while $\ket{\Omega}\equiv\left| \{\rho=1/2\} \right>$ is regarded as a zero-magnetisation ground state. \\

Hence, the quench dynamics takes place by switching off the initial potential $V_j$ at time $t=0$. Consequently, the gas expands freely to the right side of the chain and develops a non-trivial profile of density around the junction at $j=0$, which enlarges with time. More precisely, in the hydrodynamic limit $L\to \infty$,  $t \rightarrow \infty$, $ j  \rightarrow \infty $ with $t\leq L$ and $j/t$ fixed (see Subsec.~\ref{SemiClassicalHydroDescription}), a non-trivial density profile forms in the region $-t\leq j \leq t$ according to~\cite{Antal1999,Antal2008}
\begin{equation}\label{dens-intro}
\rho(j,t) \, =
\begin{cases}
	 \frac{1}{2\pi} {\rm arccos} \frac{j}{t} & \text{ if } \left| \Omega \right>= \left| \{ \rho=1/2\} \right> \\
	 \frac{1}{\pi} {\rm arccos} \frac{j}{t} &  \text{ if } \left| \Omega \right>= \left| \{ \rho=1 \} \right> 

\end{cases}
\end{equation}
at times $0<t \leq L$. 

As a result of the expansion of the gas, quantum correlations spread from the junction $j=0$ towards outer regions, leading to a growth of the entanglement with a non-homogeneous behaviour along the chain. In Refs.~\cite{Alba2014,Gruber2019,StefanoJerome} (resp.~ Refs.~\cite{Dubail2017,Rottoli2022,Ares2022}) such growth for the $n$-R\'enyi entropy, defined as 
\be
S_n(j,t)= \frac{1}{1-\alpha} \log  \tr \ (\rrho_A(t) )^n,
\ee
and its limit $n \to 1$ where it reduces to the von Neumann entanglement entropy 
\be
S_1(j,t)= - \tr \rrho_A(t) \log \rrho_A(t),
\ee 
has been thoroughly studied for the reduced density matrix of the subsystem $A= [j,+\infty]$ for the half-filled (resp. fully-filled) case.
In this work, we compute and study  the symmetry-resolved counterparts of the above quantities, that are 
\be
S_{n,N}(j,t)= \frac{1}{1-\alpha} \log  \tr \ (\rrho_A(t,N) )^n
\ee
for the symmetry-resolved $n$-R\'enyi entropy and  
\be
S_{1,N}(j,t)= - \tr \rrho_A(t,N) \log \rrho_A(t,N)
\ee 
for the von Neumann entanglement. Here, $\rrho_A(t,N)$ is defined via Eq.~\eqref{rhoN} by taking the hydrodynamic limit of the expanding quantum gas in Eq.~\eqref{model}, as  further specified below. 

%%%%%%%%%%%%%%%%%%%%%%%%%%%%%%%

\subsection{Phase-space hydrodynamics} \label{SemiClassicalHydroDescription}
As a first step towards the calculation of the symmetry-resolved entanglement during the quench dynamics, we move to a hydrodynamic description of the problem. In this way, we eventually obtain a quasi-classical description of the time evolution from which asymptotically exact result for the conserved charges readily follows, see e.g. Ref.~\cite{Antal1999,Antal2008,Karevski2002,Rigol2004,Platini2005,Platini2007}. However, this machinery is not sufficient for the description of quantum effects such as entanglement at zero temperature. In fact, the latter are captured only after the re-introduction of large-scale quantum fluctuations on top of the phase-space hydrodynamics, as detailed in the next subsection.\\

The quasi-classical treatment consists of taking an appropriate hydrodynamic limit i.e., considering the model at large space-time scales by sending $L,j, t\to\infty$ keeping fixed the ratio $j/t$. In such a limit, the model can be described in terms of fluid cells $\Delta x$ labeled by $x$ each containing a large number of particles $\Delta x=[j,j+M]$ with $M\gg 1$. It follows that the Hamiltonian \eqref{model} can be rewritten as \cite{Wendenbaum2013,StefanoJerome}
\be
\Ha=-\frac{1}{2}\int_{-L}^{\infty} \dd x \int_0^{\Delta x} \frac{\dd y}{\Delta x} \ \left(\Cc^\dagger_{x+y} \Cc_{x+y+1} + \text{h.c.}\right)
\ee
and can be diagonalised in Fourier basis within each fluid cell as
\be
\Ha= -\int_{-L}^{\infty} \dd x \ \int_{-\pi}^\pi \frac{\dd k}{2\pi} \cos k \ \hat\eta^\dagger_{k,x} \ \hat\eta_{k,x}
\ee
where  
\be
\Cc^\dagger_{x+y}=\int_{-\pi}^\pi \frac{\dd k}{2\pi} \ e^{\I k y} \ \hat\eta^\dagger_{k,x}, \qquad \hat\eta_{k,x}=(\hat\eta_{k,x}^\dagger)^\dagger.
\ee
The two variants of the initial state that we investigate fill the left part of the system with modes $-\pi\rho_0\leq k \leq \pi\rho_0$ ($0\leq \rho_0\leq 1$), leaving empty the right side. The specific choice  $\rho_0=1$ (resp.,~$\rho_0=1/2$) associated with $\left| \{\rho=1\} \right>$ ($\left| \{\rho=1/2\} \right>$) corresponds to l.h.s. of the system being entirely filled (half filled). \\
Crucially, both initial states are asymptotically described by a Wigner function which is, in our cases, equivalent to the local occupation function of the free particles and reads as
\be
W_0(x,k)=\begin{cases} 1, \quad \text{if $x\leq 0$ and $-\pi\rho_0\leq k \leq \pi\rho_0$}; \\[4pt] 0, \quad\text{otherwise.}\end{cases}
\ee
Its evolution in phase-space is dictated by the Euler equation \cite{Ruggiero2019,StefanoJerome}
\be
\de_t \ W_t(x,k) + \sin k \ \de_x  \ W_t(x,k)=0
\ee
with the simple solution
\be
W_t(x,k)=W_0(x-t\sin k,k),
\ee
see also Ref.~\cite{Fagotti2017,Fagotti2020} for details on the derivation. An important consequence of the above solution is that the dynamics at zero-temperature is characterised by the zero-entropy condition $W_t=\{0,1\}$ of the local macro-states at each time. It follows that one can focus only on the hydrodynamic evolution of local Fermi points $k^\pm_F(x,t)$, satisfying the so-called zero-entropy GHD equation \cite{Doyon2017}
\be\label{zero-entropyGHD}
\left(\de_t + \sin k^\pm_F \ \de_x\right) k^\pm_F=0.
\ee
The solution of Eq.~\eqref{zero-entropyGHD} allows us to built the Fermi contour $\Gamma_t$ as
\be\label{gamma}
\Gamma_t=\left\{ (x,k) \ : \ k_F^-(x,t)\leq k \leq k_F^+(x,t)\right\} \ee
and to re-construct the time-evolved Wigner function simply as 
\be
W_t(x,k)=\begin{cases} 1, \quad \text{if $k_F^-(x,t)\leq k \leq k_F^+(x,t)$}; \\[4pt]  0, \quad \text{otherwise.}\end{cases}
\ee 
Notice that the quench problem under analysis is characterised by a connected Fermi sea at each time \cite{Allegra2016,Dubail2017,Gruber2019,StefanoJerome,Scopa2021b}, i.e., it displays only two local Fermi points $k_F^-(x,t)\leq k_F^+(x,t)$ resulting in the Fermi contour of Eq.~\eqref{gamma}. The time-evolved Fermi contour $\Gamma_t$ is a key quantity for our study, not only because it fully encodes the quasi-classical dynamics of the model, but also because it constitutes the background over which quantum fluctuations are re-introduced, as shortly presented. \\

%The above reviewed semi-classical hydrodynamical treatment is, nevertheless, already very powerful method. 
As already mentioned, once the Fermi contour is determined, one has immediate access to the exact asymptotic profile of conserved charges densities $q$ and currents $j_q$ as
\begin{subequations}\be
q(x,t)=\int_{k_F^-(x,t)}^{k_F^+(x,t)} \frac{\dd k}{2\pi} \ h_q(k);
\ee
\be
j_q(x,t)=\int_{k_F^-(x,t)}^{k_F^+(x,t)} \frac{\dd k}{2\pi} \ \sin k \  h_q(k),
\ee
\end{subequations}
where $h_q(k)$ is the single-particle eigenvalue associated to the charge $q$ (for instance: $h_1\equiv 1$ for the particle density, $h_2\equiv -\cos k$ for the energy density and so on).\\

For sake of concreteness, in the cases  $\rho_0=\{1,1/2\}$, one finds the solutions for $0\leq x/t\leq1$
\be
k_F^\pm(x,t)=\begin{cases} \left\{ \pi-\arcsin(x/t); \ \arcsin(x/t)\right\}; \quad \rho_0=1
\\[4pt]
\left\{\pi/2; \ \arcsin(x/t)\right\}; \qquad \rho_0=1/2
\end{cases}
\ee
for the Fermi points, and 
\be\label{density}
\rho(x,t)=\begin{cases}
(\rho_0/\pi) \arccos(x/t),\qquad \text{if $|x|/t\leq 1$}; \\[4pt]
\rho_0, \qquad \text{if $x/t<-1$};\\[4pt]
0, \qquad \text{otherwise}
\end{cases}
\ee
for the density profile. Given a bi-partition of the system as
\be\label{bi-partition}
A\cup \bar{A} \qquad \text{with $A=[-L, x]$,}
\ee
we compute, for future convenience, the number of particles in $A$ as function of the cutting point $x$ and of time $t$
\be\label{NA}
N_A(x,t)=\int_{-L}^x \dd y \ \rho(y,t)\equiv \rho_0\ {\cal N}(x,t)
\ee
with scaling function
\be\label{resc-NA}
{\cal N}(x,t)=\begin{cases}
L-t/\pi\sqrt{1-x^2/t^2}+(x/\pi)\arccos(x/t), \\
 \text{if $|x|/t\leq 1$};\\[8pt]
(L+x), \qquad \text{if $x/t<-1$};\\[4pt]
L, \qquad \text{otherwise}.
\end{cases}
\ee
At $x=0$, we find simply
\be
{\cal N}(0,t)=L-t/\pi.
\ee
\begin{figure}[t]
\centering
(a) \hspace{3cm} (b) \\
\includegraphics[width=\textwidth]{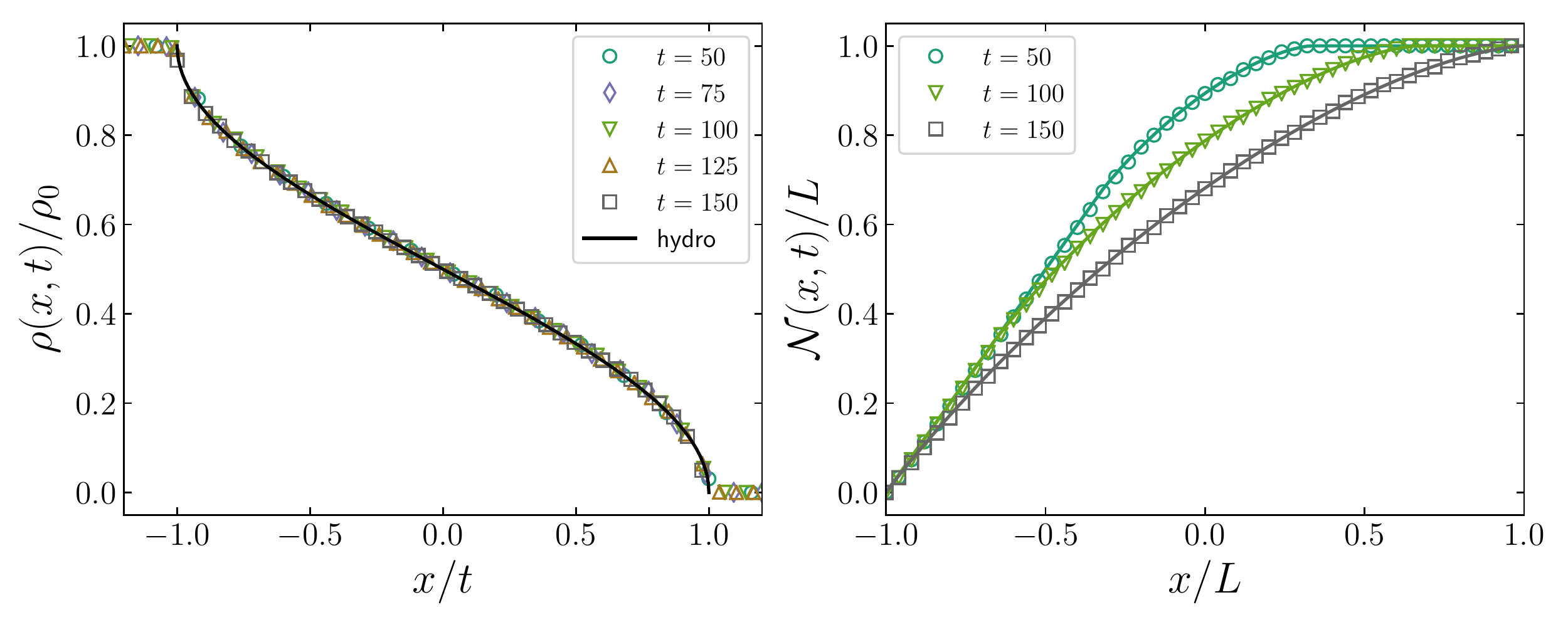}
\caption{(a) Particle density $\rho(x,t)$ in Eq.~\eqref{density} and (b) rescaled particle number ${\cal N}(x,t)$ of the subsystem $A=[-L,x]$ in Eq.~\eqref{resc-NA} as function of the rescaled position at different instants of time. The symbols are numerical data obtained for a lattice of $300$ sites while the full line is the hydrodynamic prediction.}\label{fig1}
\end{figure}
In Fig.~\ref{fig1}, the semi-classical hydrodynamic results in Eqs.~\eqref{density} and \eqref{NA} for the particle density $\rho(x,t)$ and number $N_A(x,t)$ are compared to exact numerical data obtained for the lattice model in Eq.~\eqref{model}, see~\ref{app:NUM} for details on the numerical implementation. 

\subsection{Quantum fluctuating hydrodynamics}\label{sec:QGHD}
\begin{figure}[t]
\centering
\includegraphics[width=0.7\textwidth]{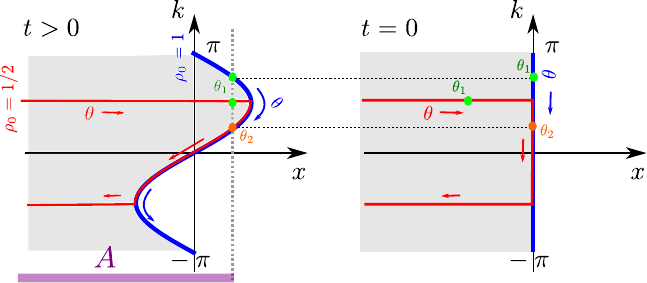}
\caption{Illustration of the Fermi contour $\Gamma_t$ at $t>0$ (left panel) and at $t=0$ (right panel) for the cases $\rho_0=1/2,1$. It is shown that a bi-partition $A=[-L,x]$ at a time $t>0$ can be encoded by the coordinates $\theta_{1,2}$ along the Fermi contour, which are then mapped backward in time to the initial Fermi contour where they can be more easily parametrised, see Ref.~\cite{StefanoJerome} for details.}\label{fig:illustration-cut}
\end{figure}

As we above mentioned, for the calculation of the entanglement entropy it is essential to restore the quantum fluctuations on top of the semi-classical hydrodynamic solution that we previously determined\cite{Ruggiero2019,Ruggiero2020,Collura2020,StefanoJerome,Ruggiero2022}. A useful and successful way to do so is to incorporate only those quantum processes that are relevant at low-energy, which can be described in terms of a Luttinger liquid. Therefore, we introduce a large-scale density fluctuation field as
\be
\delta\hat\rho=\frac{1}{2\pi} \de_x \hat\varphi
\ee
and we expand the time-dependent fermionic operators in terms of the low-energy fields of the underlying Luttinger liquid
\be
\begin{matrix}
&\Cc^\dagger_x(t) \propto \exp\left[\frac{\I}{2}\left( \hat\varphi_+ - \hat\varphi_-\right)\right] + \dots \\[4pt]
&\Cc_x(t) \propto \exp\left[\frac{\I}{2}\left( \hat\varphi_- - \hat\varphi_+\right)\right] + \dots
\end{matrix}
\ee
retaining only the leading order terms, i.e., those with smallest scaling dimensions. The above identification is valid up to a non-universal amplitude and to a semi-classical phase that are unimportant for our scopes. It is customary and useful to denote the chiral components of $\hat\varphi$ as $\hat\varphi=\hat\varphi_+ +\hat\varphi_-$. The dynamics of these quantum fluctuations is then established by the following effective Hamiltonian \cite{Dubail2017,Brun2017,Brun2018, Ruggiero2019,Scopa2020,Bastianello2020, Ruggiero2020,StefanoJerome,Scopa2021b,Ares2022}
\be\label{LL}
\Ha[\Gamma_t]=\int_{\Gamma_t} \frac{\dd\theta}{2\pi} {\cal J}(\theta)\ \sin(k(\theta)) \ (\de_\theta \hat\varphi_a)^2
\ee
together with the parametrisation of the Fermi contour
\be
\Gamma_t=\left\{ (x(\theta),k(\theta)) \ : \ k(\theta)=k_F(x(\theta),t)\right\},
\ee
in terms of $\theta\in 2\pi \mathbb{R}/\mathbb{Z}$, which is a coordinate along the contour $\Gamma_t$, $a\equiv a(\theta)=\pm$ iff $k(\theta)\gtrless 0$ and ${\cal J}(\theta)$ is simply the Jacobian of the coordinate change. In our cases, the large-scale quantum fluctuations are obtained from the ground state of the Luttinger liquid Hamiltonian \eqref{LL} at time $t=0$. Over the course of time evolution, these fluctuations are then simply transported along the Fermi contour, which gets modified according to the semi-classical hydrodynamics of Sec.~\ref{SemiClassicalHydroDescription}. \\
Importantly, in our quench setting, any bi-partition of the system $A\cup\bar{A}$ with cut in real space at position $x$ (i.e., $A=[-L,x]$) can be identified with two boundary points $\theta_{1,2}$ along the curve $\Gamma_t$ such that $k(x(\theta_{1,2}),t)=k_F^\pm(x,t)$, see Fig.~\ref{fig:illustration-cut} for an illustration.\\

We conclude this section with a remark. Although the semi-classical hydrodynamics of Sec.~\ref{SemiClassicalHydroDescription} is found to be the same for $\rho_0=\{1,1/2\}$ (up to a simple rescaling of the profiles), the same  is not true for the behaviour of quantum fluctuations. In fact, the parametrisation of the Fermi contour $\Gamma_t$ as well as the final result for the entanglement entropy is strongly dependent on the value of $\rho_0$, see the subsequent section for a brief summary and Refs.~\cite{Dubail2017}-\cite{StefanoJerome}  for a comprehensive discussion of the two cases.
%%%%%%%%%%%%%
%%%%%%%%%%%%%%

%%%%%%%%%%%%%%%%%%%%%%%%%%%%%%%%%%%

\section{Total R\'enyi entropies and charged moments}\label{sec:charge-mom}
The quantum fluctuating hydrodynamic framework enable us to exactly determine the non-equilibrium dynamics of both the total R\'enyi entropies (first computed in Ref.~\cite{Dubail2017,StefanoJerome}) and of the symmetry-resolved charged moments (cf. Eq.~\eqref{c}) in a similar fashion, as we now discuss.\\
In the original formulation, the essence of this computation for $S_n$ is in fact the determination of the one-point function of a specific field, namely the branch-point twist field $\hat{\cal T}_n$, associated with the permutation symmetry of the $n$ copies of the Luttinger liquid \eqref{LL} in the replica approach, see e.g. Refs.~\cite{cc-04,Cardy2008,Calabrese2009} and the discussion below. %Similarly, the extension to symmetry resolution requires the replacement of $\hat{\cal T}_n$ with the recently discovered composite (branch point) twist fields $\hat{\cal T}_{n,\alpha}$ \cite{gs-18}. 
Similarly, the extension to symmetry resolution requires the replacement of $\hat{\cal T}_n$ with the so-called composite (branch point) twist fields $\hat{\cal T}_{n,\alpha}$, first introduced in Ref.~\cite{OCA11,OCA14} and recently used for the calculation of charged moments \cite{gs-18}.
This field can be regarded as the fusion of the standard branch point twist field $\hat{\cal T}_n$ with a $U(1)$ vertex field $\hat{\cal V}_\alpha$, that is, 
\be
\hat{\cal T}_{n,\alpha}=\hat{\cal T}_n \times \hat{\cal V}_\alpha\ .
\ee
The vertex operator is associated with the internal symmetry of the non-replicated model and corresponds to the insertion of the flux on one of the Riemann sheets.
Notice that in absence of flux-insertions (i.e., setting $\alpha=0$), the vertex field $\hat{\cal V}_0\equiv \mathds{1}$ and we recover the usual twist field.
With these considerations, we can relate the charged moments in Eq.~\eqref{c} to the expectation value of the composite twist field as \cite{gs-18}
\be
\label{logZ}
\begin{split}
\log Z_{n,\alpha}(x,t) &\equiv  \log\tr\left[\hat\rho_A^n\  e^{\I\alpha \hat{N}_A}\right]=\log\left[ \varepsilon(x,t)^{2\Delta_{n,\alpha}}\braket{\hat{\cal T}_{n,\alpha}(x,t)}\right] +\I \alpha N_A(x,t)\\[4pt]
&=\log\left( \varepsilon(x,t)^{2\Delta_{n,\alpha}} \left|\frac{\dd\theta}{\dd x}\right|_{\theta=\theta_1}^{\Delta_{n,\alpha}} \left|\frac{\dd\theta}{\dd x}\right|_{\theta=\theta_2}^{\Delta_{n,\alpha}} \braket{\hat\tau^+_{n,\alpha}(\theta_1)\hat\tau^-_{n,\alpha}(\theta_2)}\right) +\I \alpha N_A(x,t)\,,
\end{split}
\ee
where $\hat\tau_{n,\alpha}^\pm$ are the chiral components of the composite (or  standard if $\alpha=0$) branch point twist field $\hat{\cal T}_{n,\alpha}$ living at the boundary points $\theta_{1,2}$ of subsystem $A$ with scaling dimension
\be\label{dimensions}
\Delta_{n,\alpha}=\frac{h_n}{2}+\frac{h_\alpha}{2n} 
\ee
where 
\be
h_n=\frac{c}{12}\left(n- n^{-1}\right) , \quad h_\alpha=\frac{\alpha^2}{(2\pi)^2} 
\ee
are the scaling dimension of $\hat{\cal T}_n$ and $\hat{\cal V}_\alpha$ respectively, and the central charge $c=1$ for the free Fermi gas. The factor $ \varepsilon(x,t)$ appearing in Eq.~\eqref{logZ} is a short-distance regularisation, which also guarantees that the quantity in the r.h.s of Eq. \eqref{logZ} is dimensionless. As was already shown in Ref.~\cite{Dubail2017,StefanoJerome,Scopa2021b,Ruggiero2022,Ares2022,Rottoli2022}, the expression of $\varepsilon$ for a connected Fermi sea is
\be\label{cutoff}
 \varepsilon(x,t)=\frac{C_{n,\alpha}}{\sin \pi \rho(x,t)}\,,
\ee
where $\rho(x,t)$ is the particle density in Eq.~\eqref{density} and $C_{n,\alpha}$ is a known non-universal constant, see Ref.~\cite{Calabrese2010,Jin2004} and the discussion below.\\

Eq.~\eqref{logZ} is the building block for the calculation of the total and of the symmetry-resolved entropies. For concreteness, we report below the explicit derivation for the saturated case $\rho_0=1$. The same logic applies then to the half-filled case $\rho_0=1/2$, with some additional technicalities for which we address the interested reader to Ref.~\cite{StefanoJerome}.\\
For $\rho_0=1$, one finds that the coordinate $\theta$ along the Fermi contour can be simply written as
\be
\theta \equiv k + \pi
\ee
and, therefore, one obtains
\be
\theta_1=2\pi-\arcsin\frac{x}{t}; \qquad \theta_2=\pi+\arcsin\frac{x}{t},
\ee
for the Fermi points. The Weyl factors in Eq.~\eqref{logZ} associated with the change of coordinates read as
\be
\left|\frac{\dd\theta}{\dd x}\right|_{\theta=\theta_{1,2}}=\left(t \sqrt{1-\frac{x^2}{t^2}}\right)^{-1}
\ee
and the two-point correlation function is expressed as
\be
\braket{\hat\tau^+_{n,\alpha}(\theta_1)\hat\tau^-_{n,\alpha}(\theta_2)}=\left|2\sin\frac{\theta_1-\theta_2}{2}\right|^{-2\Delta_{n,\alpha}}=\left(2\sqrt{1-\frac{x^2}{t^2}}\right)^{-2\Delta_{n,\alpha}}.
\ee
Finally, using Eq.~\eqref{density}, we write the UV cutoff $\varepsilon$ in Eq.~\eqref{cutoff}  explicitly as
\be
\label{epsilon}
 \varepsilon(x,t)=\frac{C_{n,\alpha}}{\sqrt{1-x^2/t^2}}\,.
\ee
Putting all the elements together, one eventually obtains
\be\label{eq-L-saturated}
\log Z_{n,\alpha} =  -2\Delta_{n,\alpha} \log\left[2t \left|1-\frac{x^2}{t^2}\right|^{3/2}\right] + \I \alpha N_A(x,t)+\Upsilon_{n,\alpha},
\ee
(with $N_A$ given in \eqref{NA}) which we rewrite as
\be\label{logZ-res}
\log Z_{n,\alpha} =  -2\Delta_{n,\alpha} \log \mathcal{L} (x,t) +\I \alpha\rho_0 \ {\cal N}(x,t) + \Upsilon_{n,\alpha} \,,
\ee
in terms of the function $\mathcal{L}(x,t)$, introduced for convenience. Indeed, it is possible to show that the structure in Eq.~\eqref{logZ-res} for the charged moments holds also for the case $\rho_0=1/2$ and that the details of the specific quench protocol under consideration enters only through the definition of ${\cal L}(x,t)$. From Eq.~\eqref{eq-L-saturated} and Refs.~\cite{Dubail2017,StefanoJerome}, we find that
\be\label{L}
{\cal L}(x,t)=\begin{cases}
2t \left|1-\frac{x^2}{t^2}\right|^{3/2}; \quad \text{if $\rho_0=1$};\\[10pt]
\frac{2L}{\pi}\sqrt{|\frac{x}{t}-t(1-\frac{x^2}{t^2})|}\times\\
|\sqrt{1+\sqrt{1-\frac{x^2}{t^2}}}-\sgn(x)\sqrt{1-\sqrt{1-\frac{x^2}{t^2}}}| |\sin\frac{\pi(x-t)}{2L}|;\\[3pt]
\text{if $\rho_0=1/2$.}
\end{cases}\ee
The additive constant $\Upsilon_{n,\alpha}$ in Eq.~\eqref{logZ-res} is related to $C_{n,\alpha}$ in Eq.~\eqref{epsilon} as 
\be
\Upsilon_{n,\alpha}\equiv 2\Delta_{n,\alpha}\log \frac{C_{n,\alpha}}{2}
\ee
and it has been analytically determined in Ref.~\cite{brc-19} exploiting the Fisher-Hartwig conjecture
\be
\Upsilon_{n,\alpha}=\frac{\I n}{2} \int_{-\infty}^\infty \dd w \ \left(\tanh(\pi w)-\tanh(\pi n w + \I \alpha/2)\right) \log\frac{\Gamma(\frac{1}{2}+\I w)}{\Gamma(\frac{1}{2}-\I w)} -2\Delta_{n,\alpha} \log(2).
\ee
It is then  customary to rewrite this constant as
\be\label{non-uni-cst}
\Upsilon_{n,\alpha}= \Upsilon_n + \alpha^2 \nu_n +e_{n,\alpha},
\ee
with
\be
\nu_n=-\frac{\log(2)}{4\pi^2 n}+\frac{\I n}{8}\int_{-\infty}^{\infty} \dd w \left(\tanh^3(\pi n w) -\tanh(\pi n w)\right) \log\frac{\Gamma(\frac{1}{2}+\I w)}{\Gamma(\frac{1}{2}-\I w)},
\ee 
and $e_{n,\alpha}= {\cal O}(\alpha^4)$ (cf.~Ref.~\cite{brc-19}), such that the first term $\Upsilon_n\equiv \Upsilon_{n,0}$ reproduces the non-universal constant obtained in Ref.~ \cite{Calabrese2010,Jin2004} in the absence of fluxes $\alpha=0$.\\

\begin{figure}[t]
\centering
(a) \hspace{3cm} (b) \\
\includegraphics[width=\textwidth]{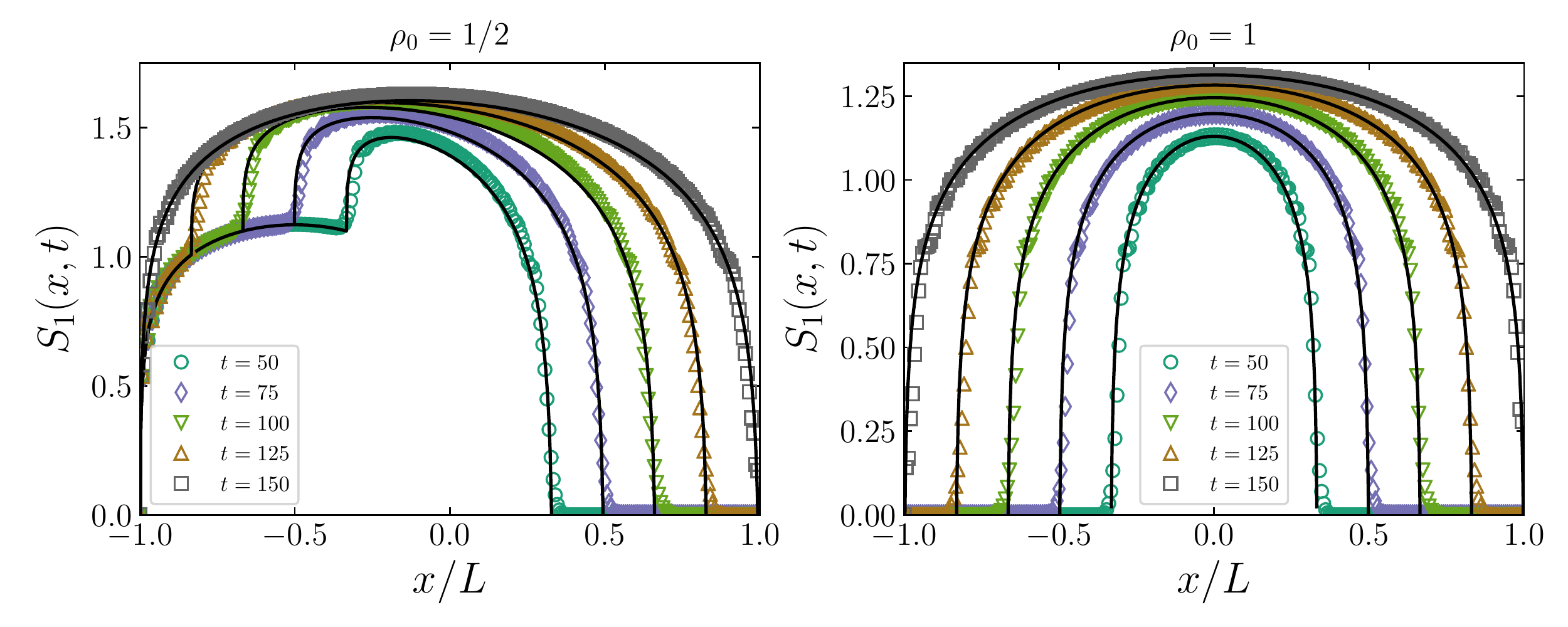}
\caption{Total von Neumann entanglement entropy for (a) $\rho_0=1/2$ and (b) $\rho_0=1$. The solid line shows the analytical prediction in Eq.~\eqref{tot-RE} provided by quantum fluctuating hydrodynamics while symbols show the numerical data obtained for a lattice of $300$ sites. Although the two settings are characterised by the same semi-classical description (up to a rescaling), cf.~Fig.~\ref{fig1}, the entanglement properties are very different. In both the cases (a)-(b), the hydrodynamic prediction is very accurate.}\label{fig:tot-EE}
\end{figure}

 The total R\'enyi entropies are recovered from  Eq.~\eqref{logZ} by plugging the correct prefactor $1/(1-n)$ and by setting the flux $\alpha=0$  \cite{Dubail2017,StefanoJerome} i.e.,
\be\label{tot-RE}\begin{split}
S_n(x,t)&=\frac{1}{1-n} \log Z_{n,\alpha}\Big\vert_{\alpha=0}=\frac{n+1}{12n}\log{\cal L}(x,t) + \frac{\Upsilon_{n}}{1-n}.
\end{split}
\ee
Notice that the constant $\Upsilon_n/(1-n)$ is related to $C_{n, 0}$ in \eqref{epsilon} as
\be
\frac{\Upsilon_n}{1-n}=-\frac{n+1}{12n}\log\frac{C_{n,0}}{2},
\ee
and gives for $n\to 1$
\be
\lim_{n\to 1} \frac{\Upsilon_n}{1-n}= \frac{\widetilde\Upsilon +\log(2)/3}{2}\,,
\ee
where $\widetilde\Upsilon\approx 0.49502$ is the Korepin-Jin constant \cite{Jin2004}, consistently with the known results for the total entropy \cite{StefanoJerome,Scopa2021b,Eisler2021}. In Fig.~\ref{fig:tot-EE}, we show the exact numerical results for the total von Neumann entanglement entropy alongside with the hydrodynamic formula in Eq.~\eqref{tot-RE} for the cases $\rho_0=1/2,1$ respectively. The agreement of the hydrodynamic prediction with the data is remarkably good.

%%%%%%%%%%%%%%%%%%%%%%%%%%%%%%%%%%%
\section{Symmetry-resolved R\'enyi entropies}\label{sec:SRRE}
\begin{figure}[t]
\centering
\includegraphics[width=0.8\textwidth]{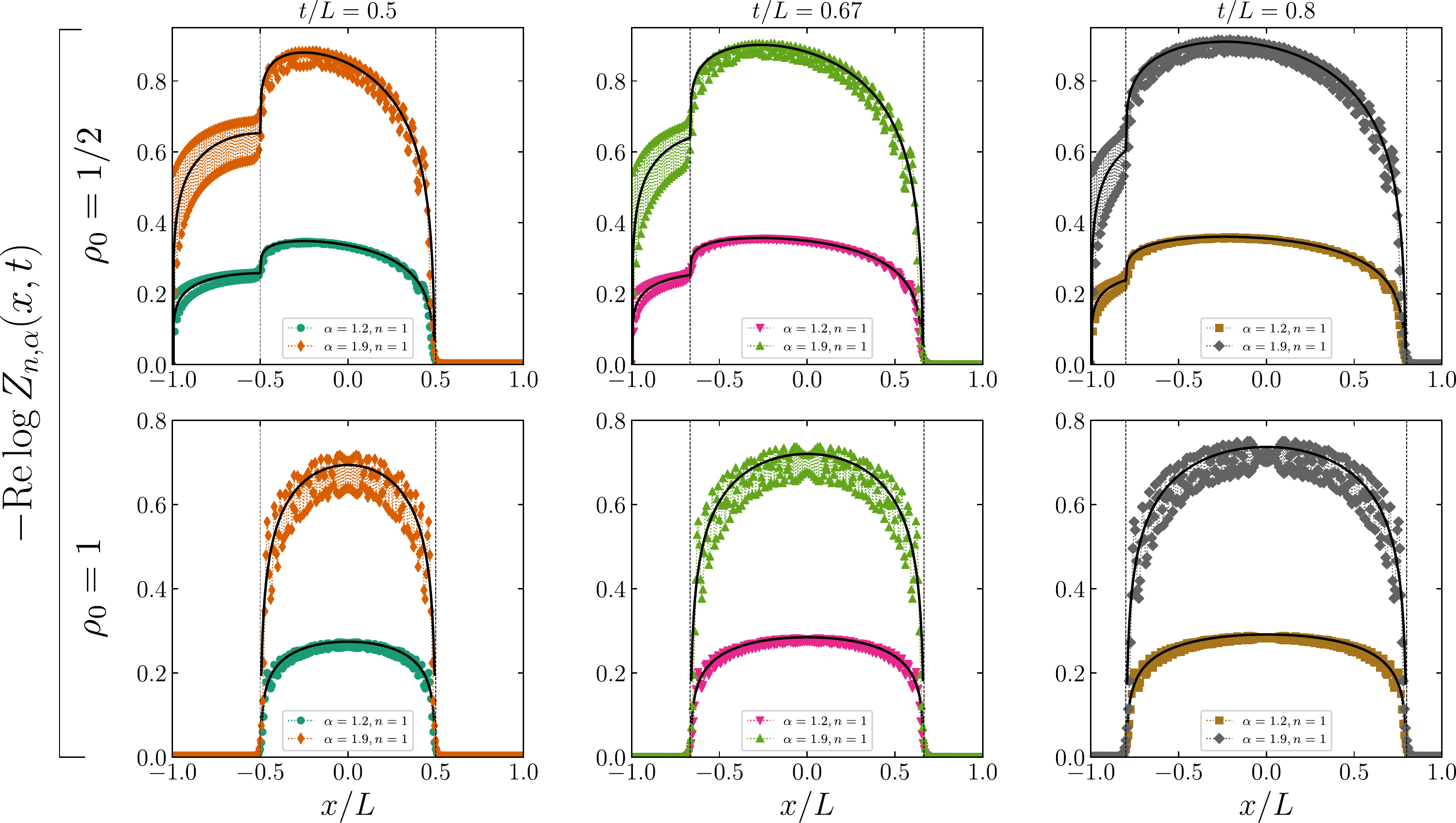}
\caption{Real part of the charged moments in Eq.~\eqref{re-mom} as function of the cutting position $x$ for $n=1$, different values of $\alpha$ (see plots legend) and at different times $t/L=0.5,0.67,0.8$ from the left to rightmost panel. The top (bottom) row shows the results for a half (fully) filled initial state $\rho_0=1/2$ ($\rho_0=1$). In each plot, the symbols show the numerical data obtained for a lattice of $300$ sites while the solid line is the analytical prediction in \eqref{re-mom}; the vertical axes mark the light cone position $|x|=t$.}\label{fig:charged-mom}
\end{figure}

\begin{figure}[t!]
\centering
(a) \hspace{5cm} (b) \\
\includegraphics[width=0.8\textwidth]{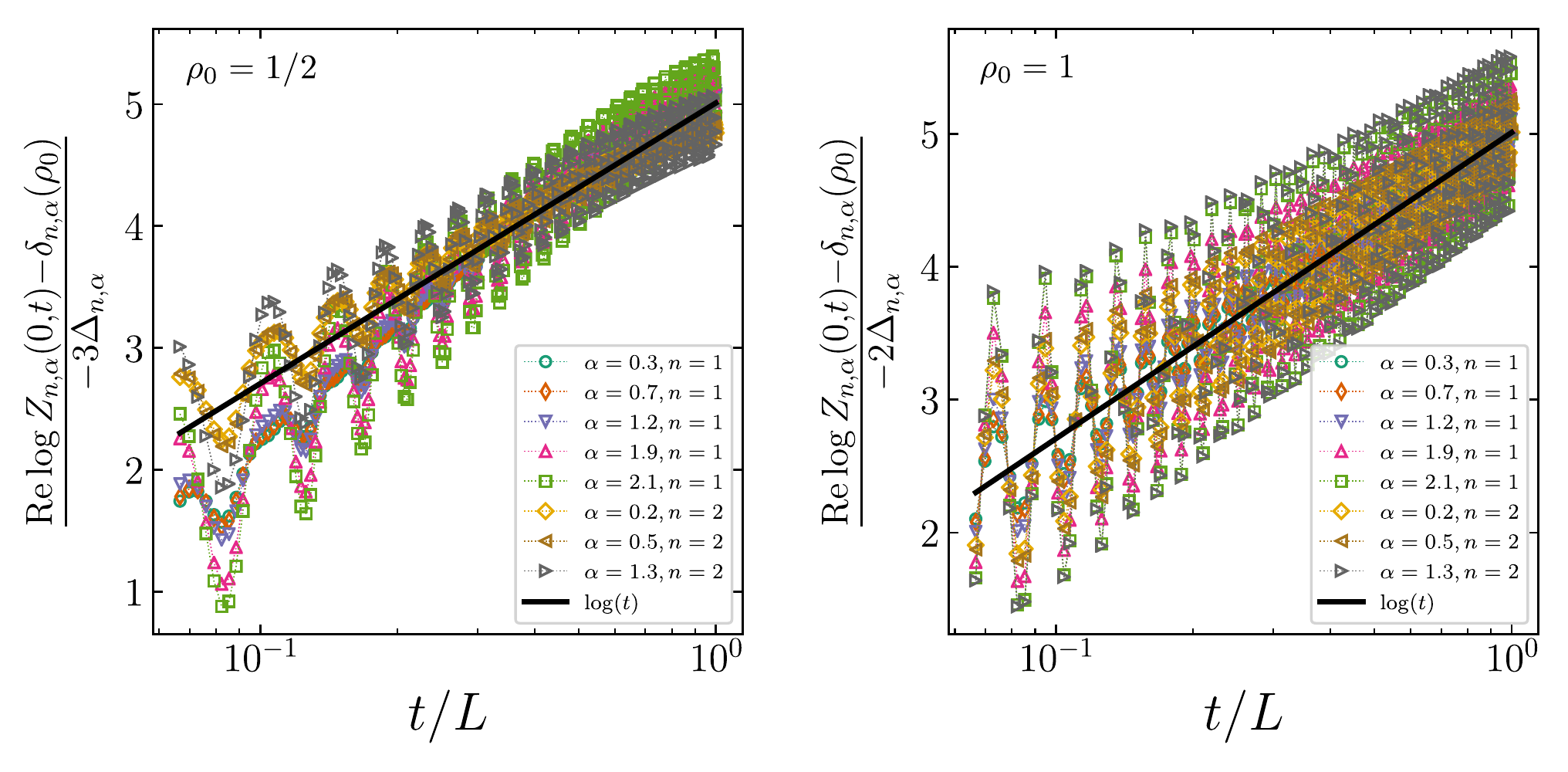}
\caption{Half system behaviour of the real part of the charged moments as function of time for different values of $n$ and $\alpha$ (see plots legend) for (a) half-filled initial state $\rho_0=1/2$; (b) fully-filled initial state $\rho_0=1$. The analytical prediction in Eq.~\eqref{exp-ReLogZ} ({\it thick solid line}) is compared with exact lattice calculations ({\it symbols}) obtained for a lattice of $300$ sites.}\label{fig:charged-mom-2}
\end{figure}

\begin{figure}[t]
\centering
\includegraphics[width=0.8\textwidth]{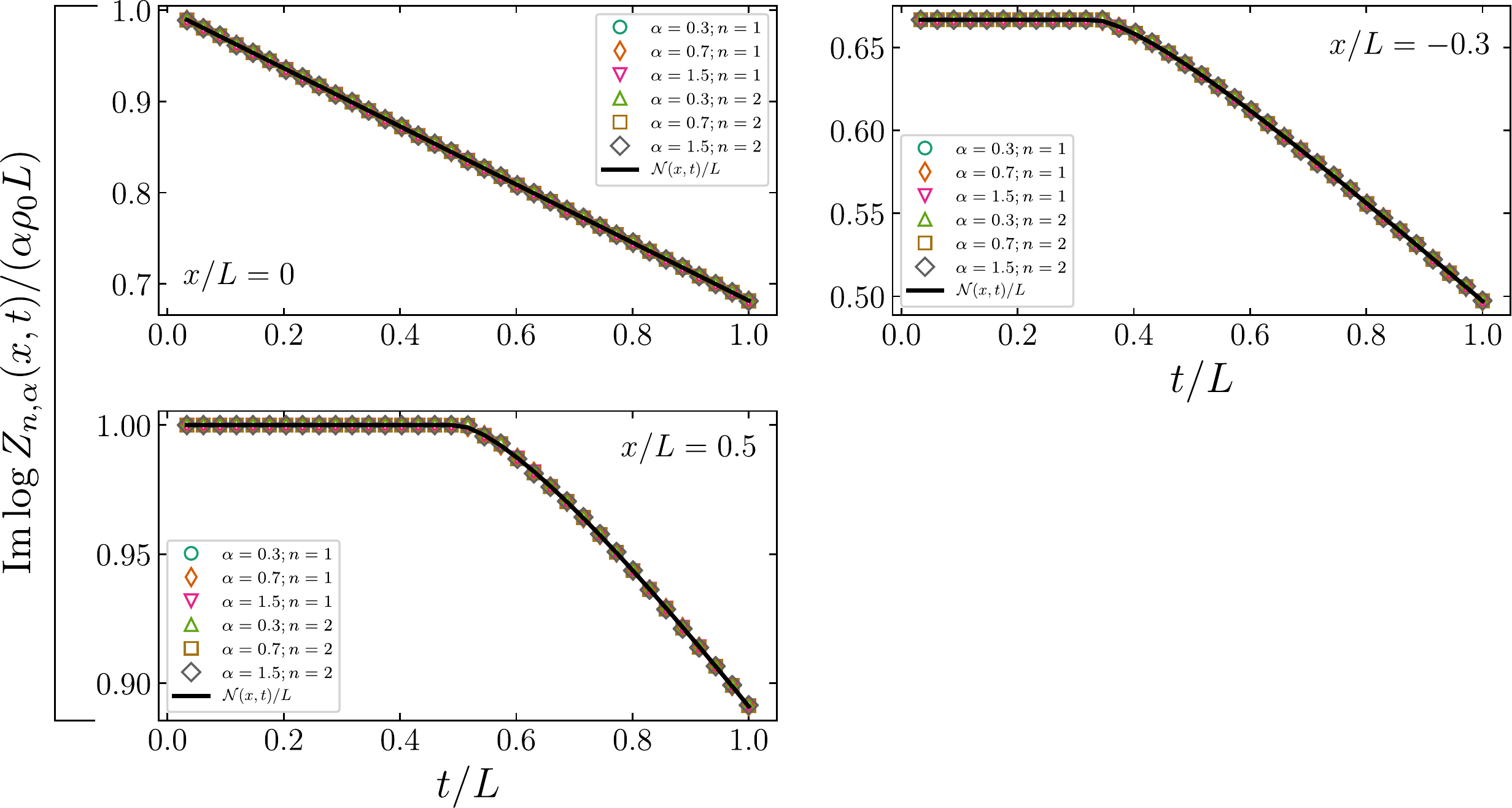}
\caption{Imaginary part of the charged moments as function of time for different cutting position $x$ (panels) and values of $\alpha$, $n$ (see plots legend). The hydrodynamic prediction in Eq.~\eqref{exp-ImLogZ} is found in agreement with exact numerical data obtained for a system of size $300$.}\label{fig:charged-mom-3}
\end{figure}

We now move towards the calculation of the symmetry-resolved R\'enyi entropies, starting from the charged moments that we computed in the previous section.\\
First, let us write down explicitly the real part of the charged moments in Eq.~\eqref{logZ-res}
\be\label{re-mom}
{\rm Re}\log Z_{n,\alpha}(x,t)=-2\Delta_{n,\alpha} \log{\cal L}(x,t)+ \Upsilon_{n,\alpha}
\ee
which at half system and for large times becomes
\be\label{exp-ReLogZ}
{\rm Re}\log Z_{n,\alpha}(0,t)\sim -2(2-\rho_0)\Delta_{n,\alpha} \log t + \delta_{n,\alpha}(\rho_0)
\ee
with $\delta_{n,\alpha}(\rho_0)\equiv\Upsilon_{n,\alpha}-2\Delta_{n,\alpha} \log(2^{\rho_0})$, and it displays a logarithmic growth for both  $\rho_0=\{1,1/2\}$, see Fig.~\ref{fig:charged-mom-2}. The imaginary part reads instead
\be
{\rm Im}\log Z_{n,\alpha}(x,t)= \alpha \rho_0 \ {\cal N}(x,t)\,,
\ee
 and at half-system it decreases linearly in time  
\be\label{exp-ImLogZ}
{\rm Im}\log Z_{n,\alpha}(0,t)= \alpha\rho_0 \left(L-t/\pi\right).
\ee
In Figs.~\ref{fig:charged-mom}-\ref{fig:charged-mom-3}, the above predictions for $Z_{n,\alpha}$ given by quantum fluctuating hydrodynamics are tested against exact lattice calculations. In particular, Fig.~\ref{fig:charged-mom} shows the real part of the charged moments as function of $x$ at different $t$ and $\alpha$ while Fig.~\ref{fig:charged-mom-2} is an analysis of the logarithmic growth in Eq.~\eqref{exp-ReLogZ} observed at half system. Finally, Fig.~\ref{fig:charged-mom-3} contains the result for the imaginary part of $Z_{n,\alpha}$. In all cases, the hydrodynamic results are found in a very good agreement.
%%%%%%%%%%
%%%%%%%%%%
%%%%%%%%%%
%%%%%%%%%%
\subsection{Fourier transform of the charged moments}
\begin{figure}[t]
\centering
\includegraphics[width=0.8\textwidth]{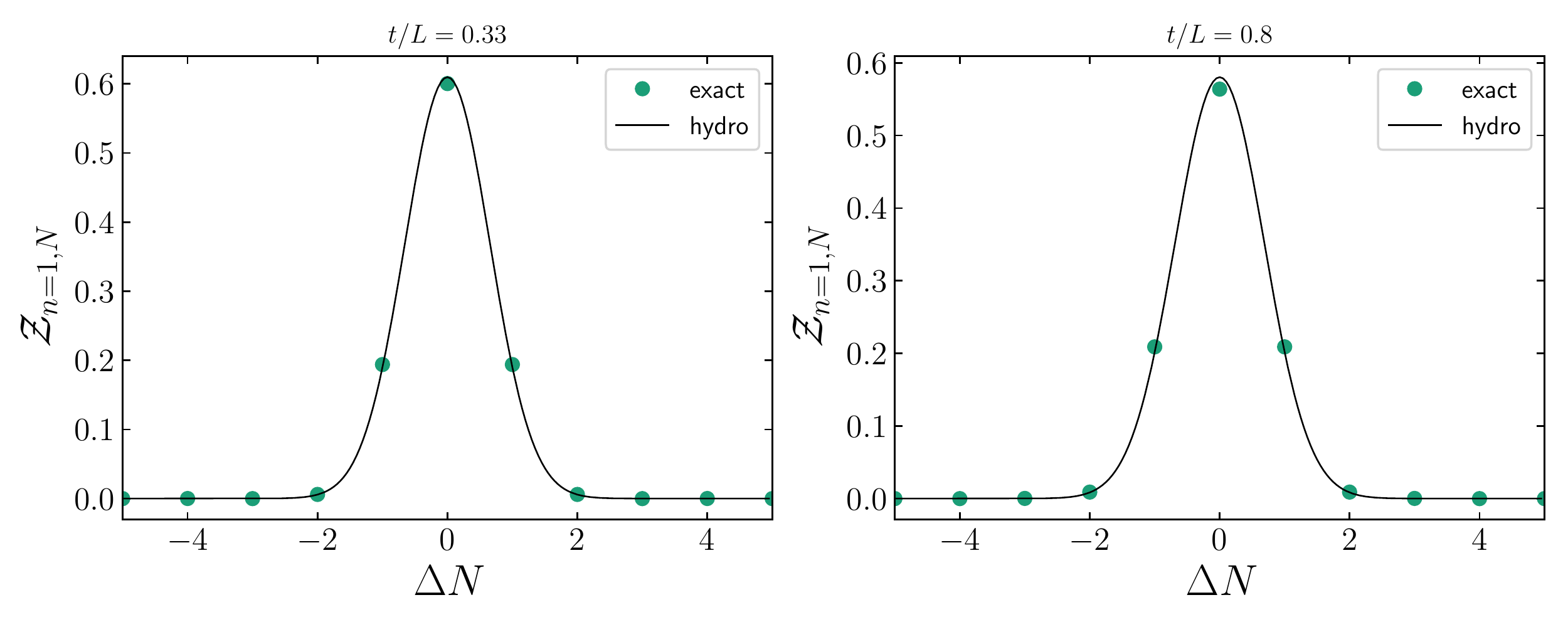}
\caption{Symmetry-resolved partition function in Eq.~\eqref{ZnN} at half system ($x=0$) as function of $\Delta N$, at $n=1$ and at different times $t/L=0.33,0.8$ (see plots legend). In each plot, the symbols show the numerical data obtained for a lattice of $300$ sites while the solid line is the analytical prediction in \eqref{Zfn}.}\label{fig:Zfn-2}
\end{figure}

\begin{figure}[t]
\centering
\includegraphics[width=\textwidth]{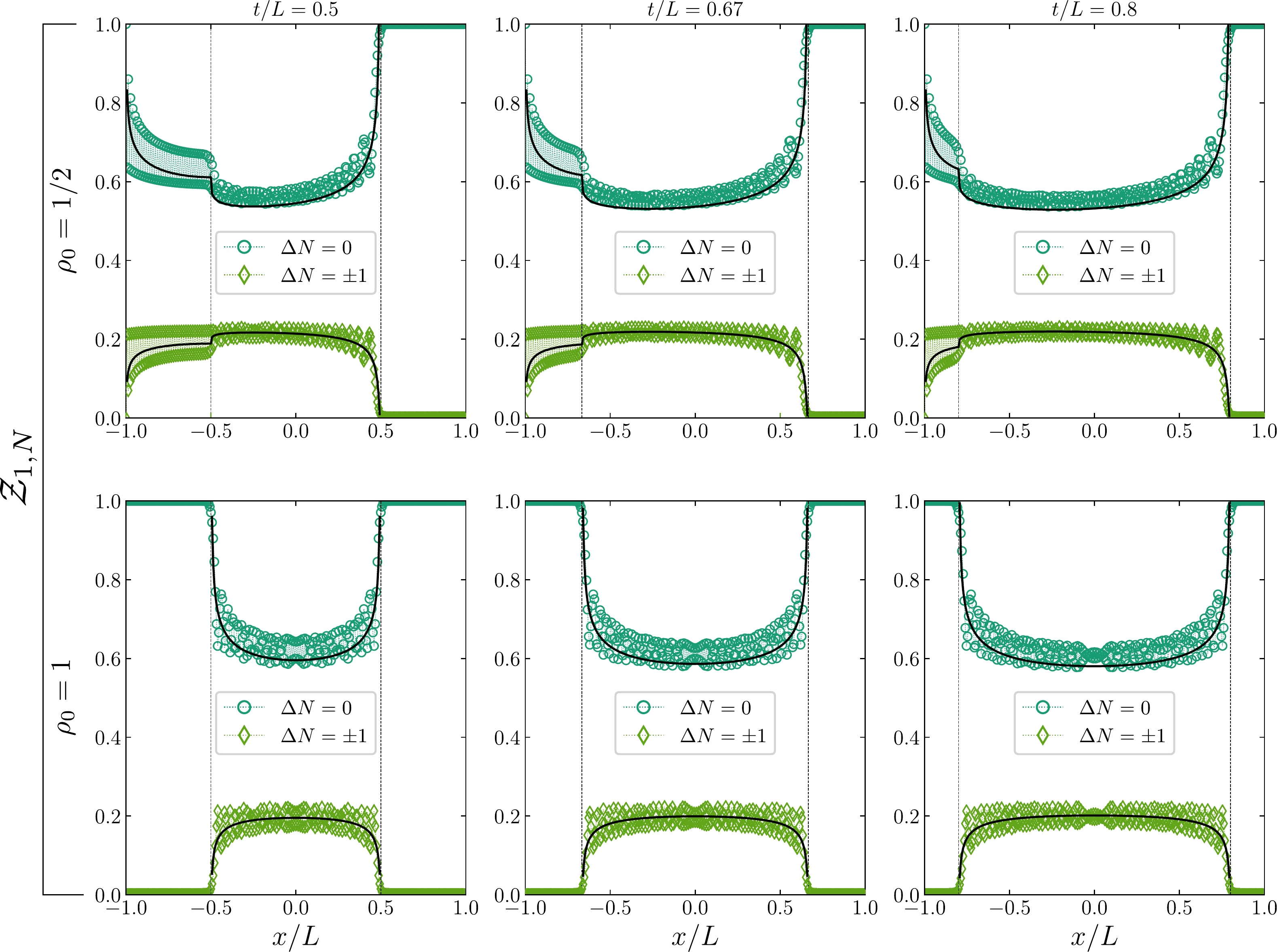}
\caption{Symmetry-resolved partition function in Eq.~\eqref{ZnN} as function of the cutting position $x$ for $n=1$, different values of $\Delta N=0,\pm 1$ (see plots legend) and at different times $t/L=0.5,0.67,0.8$ from the left to rightmost panel. The top (bottom) row shows the results for a half (fully) filled initial state $\rho_0=1/2$ ($\rho_0=1$). In each plot, the symbols show the numerical data obtained for a lattice of $300$ sites while the solid line is the analytical prediction in \eqref{Zfn}; the vertical axes mark the light cone position $|x|=t$.}\label{fig:Zfn}
\end{figure}

The next step is to compute the Fourier transform of the charged moments yielding the symmetry-resolved partition function \eqref{SRPartitionFunction}, which we denote by $\mathcal{Z}_{n,N}(x,t)$ to stress the space-time dependence.
When explicitly written, one obtains for ${\cal Z}_{n,N}$
\be\label{ZnN}\begin{split}
{\cal Z}_{n,N}(x,t)&=\int_{-\pi}^\pi \frac{\dd\alpha}{2\pi} \ e^{-\I N \alpha} Z_{n,\alpha}(x,t)\\
&=Z_{n,0}(x,t) \int_{-\pi}^\pi \frac{\dd\alpha}{2\pi} e^{-\I\alpha(N-N_A(x,t))} e^{-b_n(x,t) \alpha^2 + e(n,\alpha)}
\end{split}\ee
where 
\be\label{b-def}
b_n(x,t)=\frac{1}{4\pi^2 n} \log{\cal L}(x,t) - \nu_n
\ee
and
\be
Z_{n,0}(x,t)=\exp\left(-h_n \log{\cal L}(x,t) + \Upsilon_n\right).
\ee
The computation of the integral in Eq.~\eqref{ZnN} is performed using the saddle-point (in our case equal to a quadratic order) approximation. In particular, this amounts to ignore the contribution of $\exp(e_{n,\alpha})$ (since $\exp(e_{n,\alpha})=1+{\cal O}(\alpha^4)$, cf. Ref. \cite{brc-19}) and hence obtaining
\be\label{Zfn}
{\cal Z}_{n,N}(x,t)= \frac{Z_{n,0}(x,t)}{\sqrt{4\pi b_n(x,t)}} \exp\left(-\frac{(N-N_A(x,t))^2}{4 b_n(x,t)}\right).
\ee
It is useful to comment on the validity of this approximation, for which one can use the analogous argument of Ref. \cite{SGSRE}. In our case, one can identify the small parameter $\varepsilon$ of Ref.~\cite{SGSRE} with the inverse of $ b_n(x,t)$, which indeed becomes small as time progresses. The comparison to  \cite{SGSRE} tells us that neglecting $\exp(e_{n,\alpha})$ is legitimate if $ b_n(x,t)\gg1$ and more importantly,
\be
	(N-N_A(x,t))^2 \ll b_n(x,t)\,,
\ee
that is,  $(N-N_A(x,t))^2$ does not have to be large but can take small values if $ b_n(x,t)$ is large. In Fig.~\ref{fig:Zfn-2}, we show the probabilities ${\cal Z}_{1,N}$ at half system as function of the charge imbalance $\Delta N\equiv N-N_A(x,t)$ at different times. Notice that, due to particle number conservation, the width $b_n(x,t)$ of the Gaussian in \eqref{Zfn} is really small
(for instance, the variance at half-system for large times is $b_n(0,t) \propto {\cal L}(0,t)\sim \log(t)$) and therefore the fluctuations with $|\Delta N| \gtrsim  2$ particles are strongly suppressed.
In Fig.~\ref{fig:Zfn}, ${\cal Z}_{1,N}$ is visualised as function of $x$ for different choices of $N$ and $t$,  and compared with exact lattice numerics. 
%%%%%%%%%%
%%%%%%%%%%
%%%%%%%%%%
%%%%%%%%%%%%%%

%%%%%%%%%%%%%%%%%%%%%%%%%%%%%%%%%%%%%
\subsection{Symmetry-resolved R\'enyi entropies}

Finally, from the symmetry-resolved partition function in Eq.~\eqref{Zfn}, the symmetry-resolved R\'enyi entropies are straightforwardly obtained as (cf.~Eq.~\eqref{SRRE2})
\be\begin{split}\label{SRRE-fin}
S_{n,N}(x,t)&\equiv\frac{1}{1-n}\log\left[\frac{{\cal Z}_{n,N}(x,t)}{({\cal Z}_{1,N}(x,t))^n}\right]=S_n(x,t)+ \frac{1}{1-n} \left[\frac{n(N-N_A(x,t))^2}{4b_1(x,t)}-\frac{(N-N_A(x,t))^2}{4b_n(x,t)}\right]\\[4pt]
&\qquad -\log(2\sqrt{\pi})+\frac{1}{1-n}\log\left[\frac{b_1(x,t)^{n/2}}{b_n(x,t)^{1/2}}\right].
\end{split}\ee
At this point, we wish to consider the analytic continuation of $S_{n,N}$  in \eqref{SRRE-fin} for $n\to1$ and obtain a closed expression for the symmetry-resolved von Neumann entropy. To this end, we write the symmetry-resolved  von Neumann entropy as $S_{1,N}=-\de_n \left[\mathcal{Z}_{n,N}/\mathcal{Z}_{1,N}^{n}\right]\Big\vert_{n=1}$ and differentiate the quantity $-\mathcal{Z}_{n,N}/\mathcal{Z}_{1,N}^{n}$ with respect to $n$, exploiting the product form $\mathcal{Z}_{n,N}\equiv {Z}_{n,0}\times g_{n,N}$
with
\begin{equation}
g_{n,N}(x,t)=\frac{\exp\left(-\frac{(N-N_{A}(x,t))^{2}}{4b_{n}(x,t)}\right)}{\sqrt{4\pi b_n(x,t)}} \ .
\end{equation}
Doing so, the analytic continuation of the symmetry-resolved von Neumann entropy takes the form
\begin{equation}
\begin{split}S_{1,N}(x,t) & =\log {Z}_{1,0}+\log g_{1,N}-\frac{g_{n,N}\ \partial_{n}{Z}_{n,0}+{Z}_{1,0} \ \partial_{n}g_{n,N}}{{Z}_{1,0} \ g_{1,N}}\\[5pt]
 & =S_{1}(x,t)+\log g_{1,N}(x,t)-\frac{\partial_{n}g_{1,N}(x,t)}{g_{1,N}(x,t)}
\end{split}
\end{equation}
since ${Z}_{1,0}\equiv 1$ by construction, even in the non-homogeneous quench setting under analysis. This means that the symmetry-resolved von Neumann entropy can eventually be expressed as
\begin{equation}
\label{SRvNFinal}
\begin{split}  S_{1,N}(x,t) &=  S_{1}(x,t)-\frac{(N-N_{A}(x,t))^{2}}{4b_{1}(x,t)}-\log\left[(4\pi b_{1}(x,t))^{1/2}\right]\\[4pt]
 & -b'_{1}(x,t) \frac{(N-N_{A}(x,t))^{2}-2b_{1}(x,t)}{4b_{1}(x,t)^{2}},
\end{split}
\end{equation}
where $b'_{1}(t,x)$ denotes the derivative of $b_n$ with respect to $n$ evaluated at $n=1$, i.e.,
\begin{equation}
b'_{1}(x,t)\equiv \de_n b_n(x,t) \Big\vert_{n=1} =-(b{}_{1}(x,t)+\nu_1+\nu'_1)\,,
\end{equation}
following directly from the definition of $b_n$ in \eqref{b-def}. We recall that $\nu_n$ is related to the non-universal constant appearing in Eq.~\eqref{non-uni-cst}.  In Fig.~9, the result for the symmetry-resolved von Neumann entropy in Eq.~(76) is compared with exact lattice numerics. As for the charged moments in Fig.~4, we observe oscillations around the hydrodynamic result. In Ref.~[40, 56], similar oscillations were detected and attributed to leading correction in the subsystem size. Including such subleading effects in our approach is non trivial and goes beyond our scopes. Nevertheless, it is quite remarkable that quantum generalised hydrodynamics is able to predict the functional behaviour of the symmetry-resolved quantities modulo oscillations.\\
%%%%%%%%%%
\begin{figure}[t]
\centering
\includegraphics[width=0.8\textwidth]{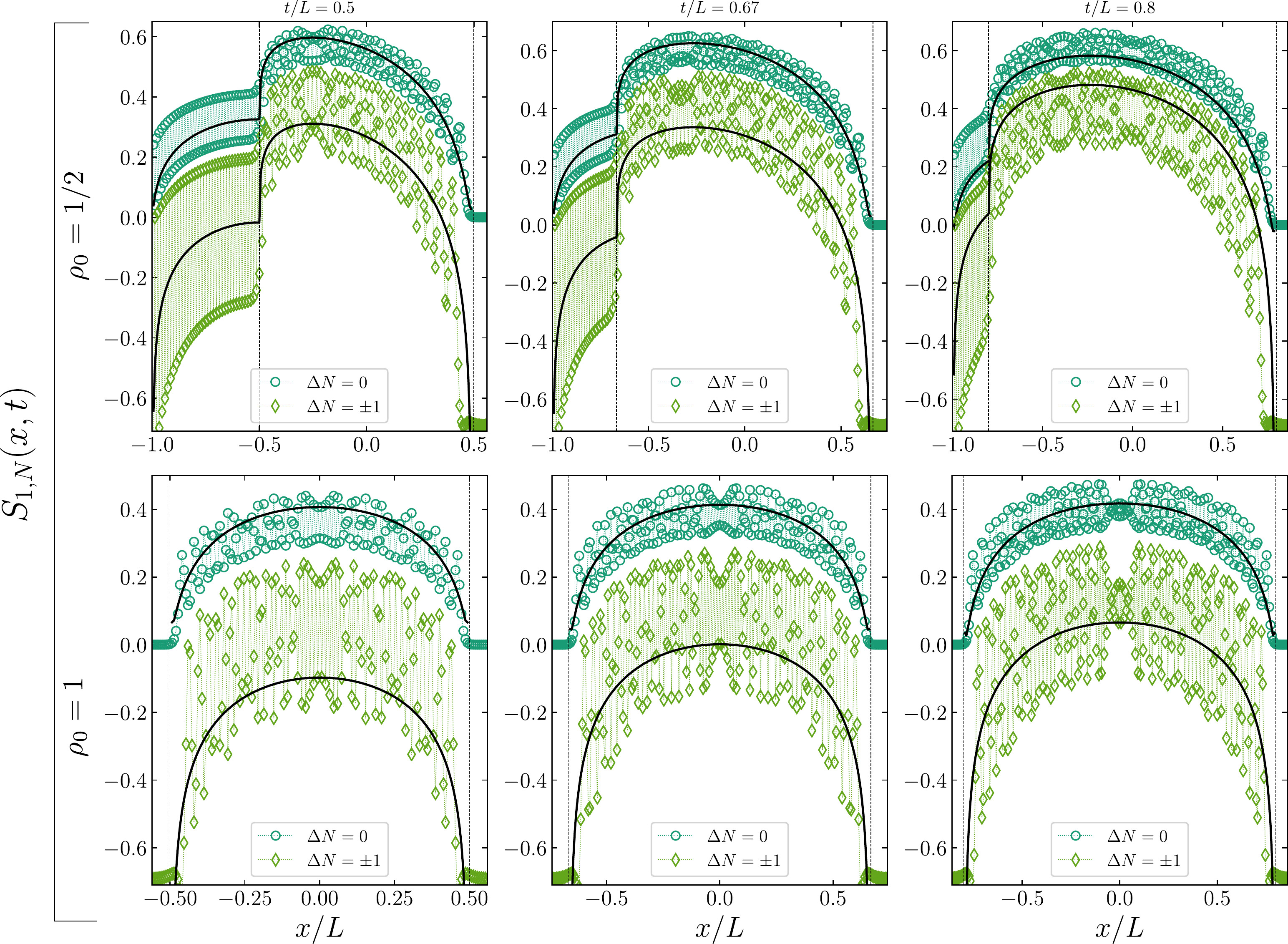}
\caption{Symmetry-resolved von Neumann entropy in Eq.~\eqref{SRvNFinal} as function of the cutting position $x$ at different values of $\Delta N=0,\pm 1$ (see plots legend) and at different times $t/L=0.5,0.67,0.8$ from the left to rightmost panel. The top (bottom) row shows the results for a half (fully) filled initial state $\rho_0=1/2$ ($\rho_0=1$). In each plot, the symbols show the numerical data obtained for a lattice of $300$ sites while the solid line is the analytical prediction in \eqref{SRvNFinal}; the vertical axes mark the light cone position $|x|=t$. The additive constants are fitted with numerics at half system.}\label{fig:SRRE}
\end{figure}
%%%%%%%%%%
\begin{figure}[t]
\centering
\includegraphics[width=\textwidth]{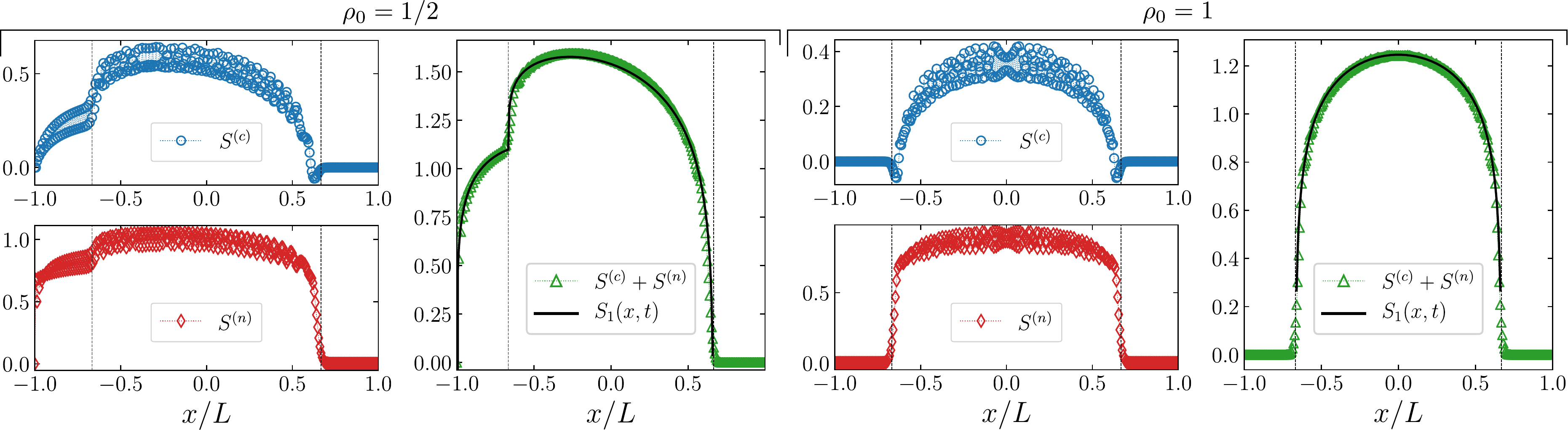}
\caption{Configurational $S^{(c)}$ and number $S^{(n)}$ entanglement entropy as function of the cutting position $x$ at time $t/L=0.67$, for different initial states $\rho_0=1,1/2$ (see plots legend). The data are obtained for a system of $300$ sites with exact lattice numerics, retaining only terms with $|\Delta N|\leq 2$ in Eq.~\eqref{Sc-Sn}. The sum of the two contributions (symbols) is found in remarkable agreement with the hydrodynamic prediction in Eq.~\eqref{tot-RE} (solid line).}\label{fig:Sc_Sn}
\end{figure}
%%%%%%%%%%%%%%%%%%%%%%%%%%%%%%%%%%%
From the analytic expressions of the symmetry-resolved von Neumann entropy in Eq.~\eqref{SRvNFinal}, one can easily investigate the limit $b_{1}(x,t)\rightarrow\infty$, which physically corresponds to a long time limit beyond the Euler scaling regime.
Expanding Eq.~\eqref{SRvNFinal} in such a $b_{1}(x,t)\rightarrow\infty$ limit, we find that
\begin{equation}
\begin{split}S_{1,N}(x,t)= & S_{1}(x,t)-\log\left[(4\pi b_{1}(x,t))^{1/2}\right]-\frac{1}{2}-\frac{2(\nu(1)+\nu'(1))}{4b_{1}(x,t)}\\
 & +\frac{\left(\nu(1)+\nu'(1)\right)(N-N_{A}(x,t))^{2}}{b_{1}(x,t)^{2}}+\mathcal{O}\left(b_{1}(x,t)^{-3}\right)\,,
\end{split}
\end{equation}
and we recall that the validity of the above expression requires that $\Delta N^2\ll b_1(x,t)$, (notice that $\nu(1)+\nu'(1)$ differs from zero). Since $b_{1}(x,t)\propto S_{1}(x,t)$, we conclude that the equipartition of entanglement in the symmetry sectors is asymptotically restored according to
\begin{equation}\label{deltaS}
\delta S_{1,N}(x,t) \sim \frac{(N-N_{A}(x,t))^{2}}{S_{1}(x,t)^{2}},
\end{equation}
with a non-trivial prefactor that depends on non-universal quantities. We finally notice that the total von Neumann entropy profile in Eq.~\eqref{tot-RE} can be recovered, for each position $x$ and time $t$, as
\be\label{Sc-Sn}
S_1(x,t)=\sum_{N}{\cal Z}_{1,N}(x,t) S_{1,N}(x,t) - \sum_{N} {\cal Z}_{1,N}(x,t)\log {\cal Z}_{1,N}(x,t)\equiv S^{(c)} + S^{(n)}.
\ee
The two terms appearing in the sum are known as configurational entanglement entropy $S^{(c)}$, measuring the total entropy due to each charge sector, and the number entropy $S^{(n)}$, which accounts for the entropy due to the charge fluctuations among different sectors, see e.g. \cite{Lukin2019}. In Fig.~\ref{fig:Sc_Sn}, we show these two contributions and we compare their sum to the hydrodynamic prediction in Eq.~\eqref{tot-RE}. Notice that the oscillations observed in $S^{(c)}$ and $S^{(n)}$ (coming from those of the charged moments, cf.~Fig.~\ref{fig:charged-mom}) nicely disappear once the two contributions are summed.

%%%%%%%%%%%%%%%%%%%%%%
\section{Summary and conclusions}\label{sec:conclusions}
We considered a one-dimensional gas of non-interacting fermions initially prepared in a bi-partite state $\ket{\Omega}$, characterised by the absence of particles on the right part $(j\geq 0)$ and by a filling on the left part  ($j<0$) of the chain with density $\rho_0=1/2$ or $1$. We subsequently let $\ket{\Omega}$ evolve unitarily with the hopping Hamiltonian in Eq.~\eqref{model} and we studied the non-equilibrium dynamics after the quench in the Euler hydrodynamic limit of large space-time scales $j,t\to \infty$ at fixed $j/t$, see Sec.~\ref{SemiClassicalHydroDescription} for details.
For this prototypical model of inhomogeneous quench setting, the non-equilibrium dynamics of conserved charges has been determined long ago (see e.g. Ref.~\cite{Antal1999,Antal2008,Karevski2002,Platini2005,Rigol2015,Rigol2004,Platini2007}) and recently complemented by results on the dynamics of the total entanglement (Ref.~\cite{Dubail2017,StefanoJerome}) and on the entanglement Hamiltonian \cite{Rottoli2022}, obtained through quantum generalised hydrodynamics. In this manuscript, we eventually completed the study on entanglement providing a careful analysis of the symmetry-resolved R\'enyi entropies as function of time and of the entangling position along the inhomogeneous system, see Sec.~\ref{sec:charge-mom} and \ref{sec:SRRE}. We found that the charged moments at half system display a logarithmic growth in time (see Eq.~\eqref{exp-ReLogZ} and Fig.~\ref{fig:charged-mom-2}) and that the symmetry-resolved von Neumann entropy is distributed among symmetry sectors with equal weights, up to corrections that scale as the inverse of the square of the total entanglement (see Eq.~\eqref{deltaS}). Our analytical results for symmetry resolved quantities are based on quantum generalised hydrodynamics and have been checked with numerical exact lattice calculations (see \ref{app:NUM} for details on the implementation), returning a very good agreement of the hydrodynamic prediction with data.
\\
Beside the per se interest of our results for the initial bi-partite state, our work aims to connect two current branches of research on entanglement, that are, symmetry resolution and quantum generalised hydrodynamics. Indeed, our discussion in Sec.~\ref{sec:charge-mom} and \ref{sec:SRRE} has general validity and can be straightforwardly extended to the study of symmetry resolved quantities in any inhomogeneous quench setting that is accessible with quantum generalised hydrodynamics, opening the doors to several subsequent analysis. For instance, it would be interesting to consider the symmetry resolution in a quartic-to-quadratic quench protocol (see e.g. \cite{Kinoshita2006}), whose total entanglement has been calculated recently in Ref.~\cite{Ruggiero2022} and realised in Ref.~\cite{Schemmer2019} with rubidium atom chips for an experimental test of the hydrodynamic results on conserved charges.
%%%%%%%%%%%%%%%%%%%%%%
\vspace{0.5cm}

{\it Acknowledgments --- } The authors acknowledge support from ERC under Consolidator grant number 771536 (NEMO). We are very thankful to Pasquale Calabrese for discussions on the project at various stages of its development and for valuable comments on the manuscript. We acknowledge Riccarda Bonsignori, Sara Murciano and Filiberto Ares for useful discussions and remarks on the manuscript.

\appendix
\section{Some details on the numerical implementation}\label{app:NUM}
In this section we consider the exact numerical calculation for free fermionic lattice Hamiltonian
\be\label{H-latt}
\Ha=-\frac{1}{2}\sum_{j=-L}^{L-1} \left(\Cc^\dagger_j \Cc_{j+1} + \Cc^\dagger_{j+1}\Cc_j \right)
\ee
which differs from that in Eq.~\eqref{model} for the presence of a right boundary. For this model, it is sufficient to determine the expression of the two-point correlation matrix 
\be
G(t)=\left[ \braket{\psi(t)|\Cc^\dagger_i \Cc_j|\psi(t)}\right]_{i,j=-L}^{L-1}
\ee
from which other quantities becomes accessible exploiting Wick's theorem. The latter is initially obtained as $G(0)=\ket{\Omega}\bra{\Omega}$ by following the state preparation discussed in Sec.~\ref{sec:intro} and subsequently evolved in the post-quench eigenstate basis $\Ha \ket{w_n}= E_n \ket{w_n}$ of the Hamiltonian \eqref{H-latt} as
\be
G(t)=\sum_{n,m=-L}^{L-1} \ket{w_n} e^{-\I t E_n} \braket{w_n|\Omega}\braket{\Omega|w_m} e^{\I t E_m} \bra{w_m},
\ee
see e.g. Ref.~\cite{Alba2014,StefanoJerome} for more details.  At this point, the density profile in Eq.~\eqref{density} is given by 
\be
\rho(i,t)= G_{i,i}(t)
\ee
and the number of particles $N_A$ in a subsystem $A$ (cf.~Eq.~\eqref{NA}) is obtained with a simple numerical integration. The next step is to build the correlation matrix restricted to the sub-system $A$ as
\be
G_A(t)= \left[ G_{i,j}(t)\right]_{i,j\in A}
\ee
and eventually to compute the set of its eigenvalues $\{ \zeta_j\}_{j=1}^{\ell_A}$, with $\ell_A$ the number of sites contained in $A$. The von Neumann entanglement entropy is then obtained as \cite{p-12,Peschel1999a,Chung2001,Peschel2003,Peschel2004,Peschel2009}
\be
S_1=-\sum_{j=1}^{\ell_A} \left[\zeta_j \log\zeta_j + (1-\zeta_j) \log(1-\zeta_j)\right].
\ee
Similarly, the charged moments can be expressed in terms of the eigenvalues $\{\zeta_j\}_{j=1}^{\ell_A}$ as \cite{gs-18}
\be\label{charged-mom-num}
\log Z_{n,\alpha}=\sum_{j=1}^{\ell_A} \log\left[\zeta_j^n e^{\I\alpha} + (1-\zeta_j)^n \right].
\ee
From Eq.~\eqref{charged-mom-num}, the other quantities (such as symmetry-resolved partition functions and entropies) are obtained via simple numerical manipulations.

\section*{References}
{

}
\end{document}